\documentclass[useAMS,usenatbib]{mn2e}
\usepackage{epsfig}
\usepackage{psfig}
\usepackage{lscape}

\newcommand{\AX}{XMMU~J185330.7-012815}
\newcommand{\G}{G31.9-1.1}

\newcommand{\chisq}{$\chi^{2}$}

\title[XMM observation of \AX]{XMM-Newton confirmation of a new intermediate polar: \AX}

\author[Hui et al.]{C. Y. Hui$^{1}$\thanks{E-mail: cyhui@cnu.ac.kr},
K. Sriram$^{2}$ and C.-S. Choi$^{2}$\\
$^{1}$Department of Astronomy and Space Science, Chungnam
National University, Daejeon 305-764, Republic of Korea\\
$^{2}$International Center for Astrophysics, Korea Astronomy and
Space Science Institute, 36-1 Hwaam, Yuseong, Daejeon 305-348,\\ Republic of Korea}

\begin{document}

\date{Received 2011 March 14}

\pagerange{\pageref{firstpage}--\pageref{lastpage}} \pubyear{2011}

\maketitle

\label{firstpage}

\begin{abstract}
We report the results from a detailed spectro-imaging and temporal analysis of 
an archival XMM-Newton observation of a new intermediate polar \AX. 
Its X-ray spectrum can be well-described by a multi-temperature thermal plasma model with the 
K-lines of heavy elements clearly detected. Possible counterparts of \AX\ have been 
identified in optical and UV bands. The low value of the inferred X-ray-to-UV and 
X-ray-to-optical flux 
ratios help to safely rule out the possibility as an isolated neutron star. 
We confirm the X-ray periodicity of $\sim238$~s but, different from the previous preliminary
result, we do not find any convincing evidence of phase-shift in this observation.
We further investigate its properties through an energy-resolved temporal analysis and 
find the pulsed fraction monotonically increases with energy. 
\end{abstract}

\begin{keywords}
binaries: close --- cataclysmic variables --- stars: individual (\AX, \G) --- X-rays: stars
\end{keywords}

\section{INTRODUCTION}

\AX\ is an X-ray object that has its emission nature not yet completely confirmed. It has been detected 
in the ROSAT All-Sky Survey (RASS). 
Based on its extent inferred from the RASS data (i.e. 11~arcmin$\times$7~arcmin), 
Schaudel (2003) has suggested that \AX, which was designated as \G\ in their work, to be a candidate of supernova
remnant (SNR) in our Galaxy. 
Schaudel (2003) have further analysed the X-ray properties of \AX\ by using 
an archival ASCA observation in which \AX\ located at an off-axis angle of $\sim14$~arcmin. 
Although they found that \AX\ appears to be elongated in an ASCA GIS image, the large off-axis angle 
precludes any constraining spatial analysis.
Examining the spectral data collected by 
ASCA, Schaudel (2003) found that the X-ray spectrum of \AX\ is featureless and can be fitted with an 
absorbed power-law model with a photon index of $\Gamma=1.84^{+0.23}_{-0.21}$ (cf. Table~5.2 in Schaudel 
2003). Together with the apparently centrally-brightened X-ray morphology, the author claimed that the power-law 
spectral fit strongly supports the interpretation of a center-filled SNR or a Crab-like SNR. However, 
with only $\sim1100$ source counts from the ASCA data, one cannot unambiguously distinguish between 
the power-law model and a single-temperature thermal plasma model with $kT=5.23^{+1.88}_{-1.35}$~keV 
(Schaudel 2003). Also, the non-detection of any radio emission from the position of \AX\ makes the 
SNR interpretation questionable. 

In an archival search for the Galactic magnetars, Muno et al. (2008) have made use of 
506 archival Chandra data and 441 archival XMM-Newton data. This search has included a dedicated 
XMM-Newton observation of \AX\ with an off-axis angle of only 
$\sim0.4$~arcmin. Interestingly, the authors have detected a signal with a period of $\sim238$~s from 
this observation. With this discovery, instead of being a magnetar candidate, Muno et al. (2008) have suggested  
this source is probably an accreting white dwarf. The cataclysmic variable nature of this 
source is further supported by the optical spectroscopy performed by J. Halpern \& E. Gotthelf 
(private communication reported in Muno et al. 2008).

Although Muno et al. (2008) have identified \AX\ as a promising candidate of cataclysmic variable, they 
have not further analysed and discussed the nature of this object as this is out of the scope of their work. 
To confirm the X-ray emission properties of \AX, a detailed spectro-imaging and temporal analysis of the 
aforementioned XMM-Newton observation is required and this provides the motivation of this investigation. 
In \S2, we are going to describe the details of this XMM-Newton observation of \AX\ as well as the procedure 
of the data reduction. The method and the results of the data analysis are presented in \S3. Finally, we will 
discuss the implication and the possible nature of \AX\ as an accreting white dwarf. 

\section{OBSERVATION \& DATA REDUCTION}

 \AX\ was observed by XMM-Newton on 25-26 October 2004 (Observation ID: 0201500301). 
 The X-ray data used in this investigation were obtained with the {\bf E}uropean {\bf P}hoton
 {\bf I}maging {\bf C}amera (EPIC) on board XMM-Newton (Jansen et al.~2001). EPIC consists of
 two Metal Oxide Semiconductor (MOS1/2) CCD detectors (Turner et al.~2001) of which half of
 the beam from two of the three X-ray telescopes is reflected to. The other two halves of the
 incoming photon beams are reflected to a grating spectrometer (RGS) (den Herder et al.~2001).
 The third of the three X-ray telescopes is dedicated to expose the EPIC-PN CCD detector solely
 (Str\"uder et al.~2001). 
 The EPIC-PN CCD was operated in small-window mode with a medium filter to block optical stray light.
 This data provide imaging spectral and temporal information. All recorded events are 
 time-tagged with a temporal resolution of 5.7 ms. The MOS1/2 CCDs were setup to operate in  
 full-window mode with a medium filter in each camera. The MOS1/2 cameras provide imaging, 
 spectral and timing information, though the later with a temporal resolution of 2.6 s only. 

 The aimpoint of the satellite in this observation is RA=$18^{\rm h}53^{\rm m}29.7^{\rm s}$
 and Dec=$-01^\circ 28' 31.8''$ (J2000). With the most updated instrumental calibration, we 
 generate the event lists from the raw data obtained from all EPIC instruments 
 with the tasks {\it emproc} and {\it epproc} of the XMM Science Analysis Software 
 (XMMSAS version 9.0.0). Examining the raw data from the EPIC-PN CCD, we did not find any 
 timing anomaly which was observed in many of the XMM-Newton data sets 
 (cf. Hui \& Becker 2006 and references 
 therein). This provides us with opportunities for an accurate timing analysis. We then created 
 filtered event files for the energy range 0.2 keV to 12 keV for all EPIC instruments and selected
 only those events for which the pattern was between $0-12$ for MOS cameras and $0-4$ for the EPIC-PN
 camera. We further cleaned the data by accepting only the good times when sky background was low and
 removed all events potentially contaminated by bad pixels. After the filtering, the effective 
 exposures are found to be 19.5~ks and 13.5~ks for MOS1/2 and EPIC-PN respectively. 

Apart from the X-ray data, we have also made use of the Optical Monitor (OM, Mason et al. 2001) data 
obtained in standard imaging mode with two 
filters UVW1 (an effective wavelength of 2910 \AA ) and UVM2 (2310 \AA).
The OM data are reduced by using the standard XMMSAS {\it omichain} task. For imaging and source detection, 
the track history was created, bad pixels were removed and the resulting image was subsequently 
used for source detection.
Since individual photons
were centroid to one eighth of detector pixel by onboard electronics which 
produces a noise pattern known as {\it modulus 8}
spatial fixed pattern noise and hence this noise was overcome using 
{\it ommodmap} task. The source detection was performed by using the task
{\it omdetect} and counts were converted into magnitudes for the corresponding filters with the 
aid of the task {\it ommag}. The UV bandpass counts 
were converted into fluxes using the recipe provided by 
Alice Breeveld.\footnote{http://xmm.esac.esa.int/sas/7.0.0/watchout/Evergreen\_tips\_and\_tricks/uvflux.shtml} 

\section{DATA ANALYSIS}
\subsection{Spatial Analysis}
\begin{figure}
\centerline{\psfig{figure=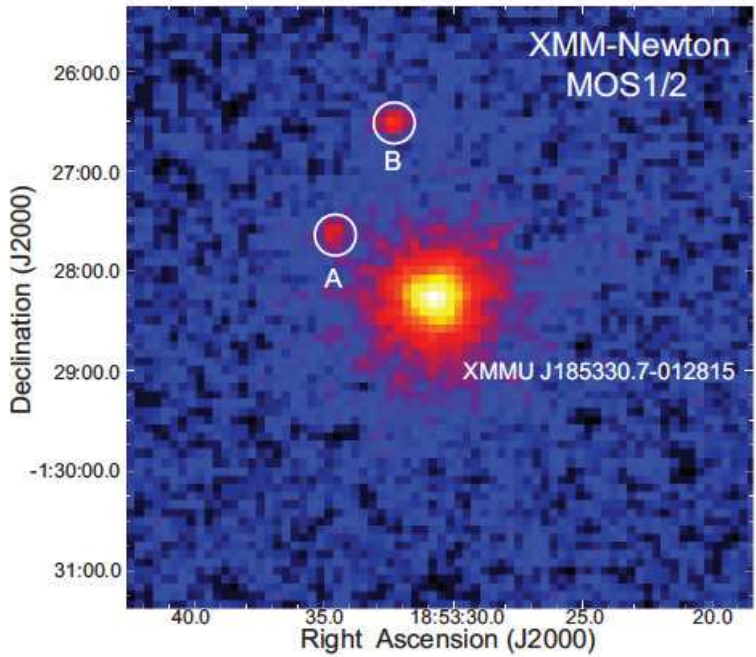,width=10cm,clip=}}
\caption[]{The raw X-ray image of the $6'\times6'$ field-of-view centered at \AX\
generated by merging the MOS1 and MOS2 data in the
energy range of $0.2-12$~keV. Two serendipitous unidentified sources are also detected in this FOV. }
\label{mos_img}
\end{figure}

The composite MOS1/2 image of a $6'\times6'$ field around \AX\ is shown in
Figure~\ref{mos_img}.  \AX\ is observed as the brightest object in this field. We determined
its position and significance by means of a wavelet detection algorithm. The X-ray position
is found to be RA=$18^{\rm h}53^{\rm m}30.708 (5)^{\rm s}$, Dec=$-01^\circ 28' 16.00 (6)''$ [J2000],
where the numbers in the parentheses indicate the $1\sigma$ statistical uncertainties of the last
digit. The signal-to-noise ratio of \AX\ is found to be $326\sigma$
\footnote{The source significances quoted in this paper are in units of Gehrels error:
$\sigma_{G}=1+\sqrt{C_{B}+0.75}$ where $C_{B}$ is the background counts.}.
This MOS1/2 image clearly rules out the 
claim of extended source as suggested by the RASS data (Schaudel 2003). 
There are two serendipitous X-ray sources found in the vicinity of \AX,
where are labeled as sources A and B in Figure~\ref{mos_img}.
The wavelet detection reports the locations and the significances of these sources to be
(RA=$18^{\rm h}53^{\rm m}34.68 (2)^{\rm s}$, Dec=$-01^\circ 27' 38.6 (4)''$
[J2000]; $S/N=6\sigma$) and
(RA=$18^{\rm h}53^{\rm m}32.24 (2)^{\rm s}$, Dec=$-01^\circ 26' 29.9 (3)''$ [J2000]; $S/N=17\sigma$)
for sources A and B respectively. Since our focus is on characterising the emission nature of \AX, 
we will not discuss the properties of these two sources further in this paper.

\subsection{Spectral Analysis}

 We estimated the effects of pileup in all EPIC data by using the XMMSAS task {\it epatplot}.
 Our results showed that all the EPIC data were not affected by CCD pileup. 
 In order to maximize the signal-to-noise ratio for \AX, we extracted its source
 spectrum from circles with a radius of 50\arcsec\ and 30\arcsec\ in the MOS1/2 and EPIC-PN cameras 
 respectively. This choice of extraction regions corresponds to the encircled energy fraction of $\sim90\%$ 
 in all cameras\footnote{http://xmm.esac.esa.int/external/xmm\_ user\_ support/documentation/uhb/node17.html} 
 and at the same time it minimizes the contamination from the nearby X-ray sources. 
 The background spectra were sampled from the nearby regions 
 with circles of a radius of 60\arcsec\ and 40\arcsec\ in MOS1/2 and EPIC-PN respectively. 
 Response files were computed for all datasets by using the XMMSAS tasks {\it rmfgen} and {\it arfgen}. 
 After the background subtraction, we have $9130\pm97$~cts, $9314\pm98$~cts, and $19282\pm140$~cts collected 
 for the spectral analysis from MOS1, MOS2 and EPIC-PN cameras respectively.

 In order to constrain the spectral parameters tightly, we fitted the data obtained from three cameras 
 simultaneously. For the spectrum extracted from each camera, we grouped the data so as to have at 
 least 50 counts per bin. We found there are fluctuations in the spectral data below 0.3~keV that can 
 be ascribed to the undesirable spectral response. Therefore, we limited all the spectral analysis 
 in the energy range $0.3-12$~keV. All the spectral fits are performed with XSPEC 12.5.1. 
 The parameters of the best-fit models are summarized in Table~1. All quoted errors are 
 $1\sigma$ for 2 parameters of interest. 

\begin{figure}
\centerline{\psfig{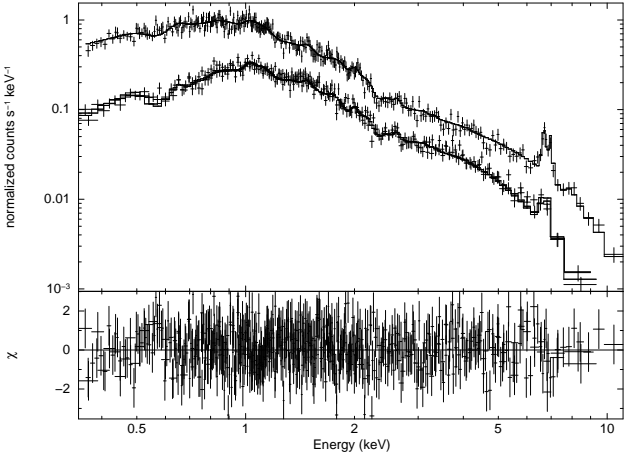}}
\caption[]{Energy spectrum of \AX\ as observed with the EPIC-PN (upper spectrum)
and MOS1/2 detectors (lower spectra) and simultaneously fitted to an absorbed
three-temperatures MEKAL model 
({\it upper panel}) and contribution to the \chisq\, fit statistic
({\it lower panel}).}
\label{spectrum}
\end{figure}

 The X-ray spectrum of \AX\ as observed by XMM-Newton is displayed in Figure~\ref{spectrum} which 
 has shown the line features suggesting K-lines of heavy elements, in particular the features at $\sim6.7$~keV 
and $\sim6.9$~keV. 
This has led us to 
 examine the spectrum with emission model of hot plasma that includes line emissions from different 
 elements. First, we attempted to fit the spectrum with MEKAL which is a code that models the plasma in 
 collisional ionization equilibrium (Mewe et al. 1985). With the metal abundances fixed at the solar values, a single 
 temperature MEKAL model results in a hydrogen equivalent column density of 
$n_{H}=(4.6\pm0.3)\times10^{20}$~cm$^{-2}$ 
 and a plasma temperature of $kT=6.4\pm0.2$~keV 
 with a goodness-of-fit of $\chi^{2}_{\nu}=1.89$ (583 D.O.F.). The large $\chi^{2}_{\nu}$ 
 indicates that this model is unlikely to be the adequate description of the data. In examining the 
 fitting residuals, we have identified the systematic deviations at energies larger than $\sim5$~keV and 
 smaller than $\sim1$~keV. This prompts us to examine the spectrum with multi-temperature plasma model.  
 
 With a two-temperature MEKAL model, the best-fit yields a similar absorption of 
 $n_{H}=4.7^{+0.3}_{-0.4}\times10^{20}$~cm$^{-2}$ and the plasma temperatures of 
 $kT_{1}=1.00^{+0.04}_{-0.03}$~keV 
 and $kT_{2}=8.34^{+0.40}_{-0.42}$~keV. In comparison with the single-temperature model, the goodness-of-fit, 
 $\chi^{2}_{\nu}=1.27$ (581 D.O.F.), is found to be improved significantly. Statistically, the additional 
 component is required at a confidence level $>99.9\%$. Although the systematic residuals 
 at energies $>5$~keV and $<1$~keV are not observed in this two-temperature model, we notice 
 that this model appears to overpredict the emission at the energies around $\sim1$~keV and $\sim6.7$~keV. 
 This leads us to speculate that either the abundances of the metals that give rise to the line emissions 
 at these energies are different from their solar values or the data require an additional component for 
 modeling. For testing the first hypothesis, we further analysed the X-ray spectrum of \AX\ 
 by taking metal abundance as free parameter in the fitting. As the residuals at these 
 energies are likely to be due to Fe K$\alpha$ and L lines, we take the abundance of iron as a fitting 
 parameter.

 By entangling the free parameters of Fe abundances in both components, we found the residuals at 
 $\sim1$~keV and $\sim6.5$~keV can be minimized. 
 The goodness-of-fit is improved ($\chi^{2}_{\nu}=1.20$, 580 D.O.F.). 
 The model yields the column density of $n_{H}=(5.7\pm0.4)\times10^{20}$~cm$^{-2}$, 
 the plasma temperatures of $kT_{1}=0.81\pm0.03$~keV and $kT_{2}=7.75^{+0.44}_{-0.65}$~keV, 
 as well as an iron abundance of Fe=$0.72\pm0.08$.  
  Although the residuals of the lines can be reduced with this model, 
 with a more careful examination, we found that there are still discrepancies between the observed data and 
 this model at energies larger than $\sim7$~keV. 
 This leads us to consider the other possibility, namely whether 
 there is any additional spectral component is required to model this data. 

 Comparison between the best-fit three-temperature plasma model and the observed data is shown 
 in Figure~\ref{spectrum}. There is no systematic deviation between the data and this model 
 within the entire adopted energy range. The goodness-of-fit is further improved ($\chi^{2}_{\nu}=1.07$, 
 579 D.O.F.). Different from the two-temperature model, we do not find any significant deviation of the Fe abundance 
 from the solar value in this adopted model. Therefore, we fixed all the metal abundances at the solar values in 
 three-temperature plasma model.  
 It results in the column density of $n_{H}=(5.5\pm0.4)\times10^{20}$~cm$^{-2}$, 
 the plasma temperatures of $kT_{1}=0.65\pm0.04$~keV, $kT_{2}=1.78^{+0.39}_{-0.15}$~keV and 
 $kT_{3}=10.84^{+1.40}_{-0.96}$~keV. The unabsorbed flux inferred by this model is 
 $f_{X}=(6.4\pm0.5)\times10^{-12}$~ergs~cm$^{-2}$~s$^{-1}$ in $0.3-12$~keV. 

 We notice that there is a class of cataclysmic variable which contains a soft blackbody component (cf. Evans \& Hellier 2007). 
 Statistically the soft blackbody component is only required at a confidence level of $\sim10\%$ in this observed spectrum. 
 Therefore, we do not consider this spectral contribution in this work. 

The ASCA and XMM-Newton spectra of \AX\ appear to be qualitatively different.
ASCA observations show no line emission features and the spectrum with a single power-law model 
 and obtained a reasonable goodness-of-fit with the ASCA data ($\chi^{2}_{\nu}=0.86$, 40 D.O.F.) (Schaudel 2003).
However, with the photon statistic improved by a factor of $\sim34$, the 
 spectral data obtained by XMM-Newton clearly show the presence of the emission line features. Fitting 
 a power-law to the EPIC spectrum does not result in an acceptable goodness-of-fit 
 ($\chi^{2}_{\nu}=1.72$, 583 D.O.F.). Similar to the single-temperature plasma model, it cannot describe the 
 data below $\sim2$~keV and above $\sim5$~keV. Such discrepancy between the inferences drawn from these two  
observations can be ascribed to the relatively poor quality of ASCA data.

\subsection{Temporal Analysis}

\subsubsection{Modulation Period}

To confirm the X-ray period by Muno et al. (2008), we did a periodicity search by using
an epoch folding method after converting the X-ray arrival times to the barycentric
times of the solar system.
The best period, which was determined by fitting a Gaussian function to the centroid of
$\chi^2$-peak, is $P=238.1\pm0.1$~s for both PN and MOS data under the assumption that the 
pulses are coherent (where the errors in $\chi^2$ 
are included and the MOS1 and MOS2 data are combined). 
This period is consistent with the result reported by Muno et al. (2008). 
If we take the possible irregularity of the pulses into account, then the error should 
increase to $P=238.1\pm1.1$~s in which the error is the standard deviation $\sigma$ of the Gaussian 
function fitted to the full $\chi^{2}$-profile. 

Muno et al. (2008) have also pointed out that the phases vary by $\approx 0.1$ cycles with 
a timescale of $\approx 5000$~s which suggest the incoherent nature of the source. 
We have also carried out the phase analysis by 
dividing the PN light curve into 4 segments with a timescale of $\approx 5000$~s
and folded each segment at the period of 238.1 s over the epoch MJD 53303.98674 (Figure~\ref{shift_check}).
By fitting a sinusoidal model to the data (see below), we obtained the following peak positions 
in time order: $1.13^{+0.02}_{-0.03}$, $1.05^{+0.06}_{-0.07}$, $1.06^{+0.03}_{-0.04}$, and
$1.14^{+0.07}_{-0.09}$ in phases (where the errors are at the 90\% confidence level).
Taking the uncertainties into account, we are not able to unambiguously state whether there is a clear, 
systematic phase shift. 
Furthermore, by running a $\chi^{2}$-test among each pair of the time-sliced light curves in 
Figure~\ref{shift_check}, their distributions  
are found to be consistent with each other at a confidence level $>99.9\%$.
Therefore, we conclude that no convincing evidence for the incoherence can be found 
in our independent analysis.

\begin{figure}
\includegraphics[angle=0,scale=0.45]{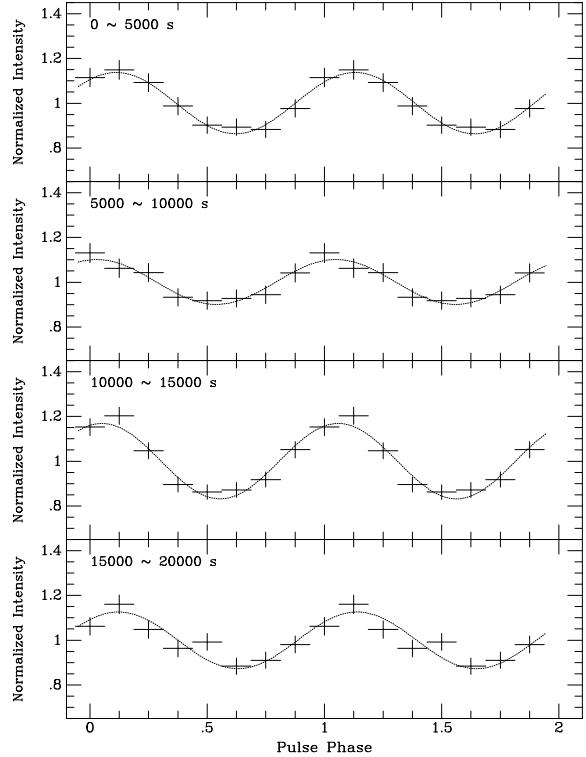}
\caption{
Pulse (or modulation) profiles for time-sliced PN data.
The data ($plus\ signs$) were folded at the period 238.1~s from the epoch MJD 53303.98674.
The pulse phase has been repeated over two cycles.
A sine curve plus a constant ($ dotted\ curve$) was fitted to the profiles.
}
\label{shift_check}
\end{figure}

\subsubsection{Energy-Resolved Folded Light Curves}

A pulse or modulation profile folded at the period of 238.1~s shows a single broad
peak over one cycle of the data.
It can be approximated by a sinusoidal model plus a constant unpulsed level:

\begin{equation}
A \sin (2\pi [\phi - \phi_0]) + C. 
\end{equation}

\begin{figure}
\includegraphics[angle=0,scale=0.45]{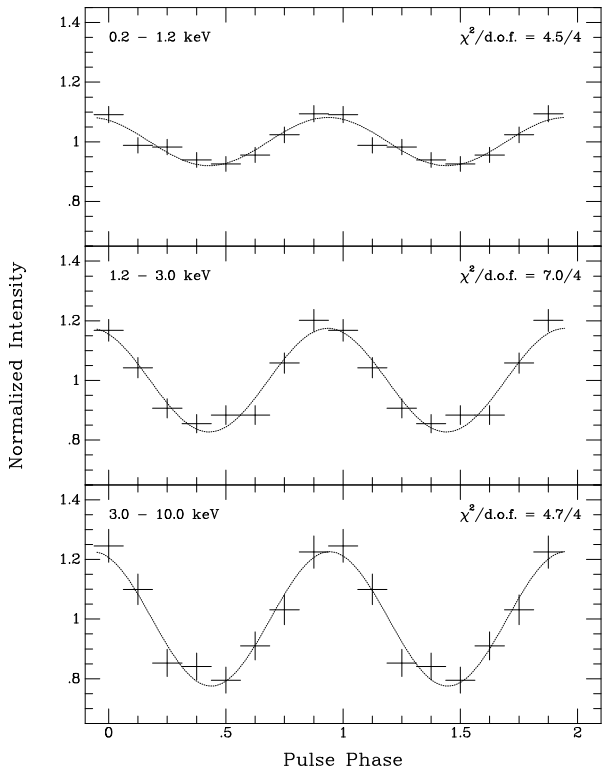}
\caption{
Energy-resolved pulse profiles of \AX\ for PN data.
The data were folded at the period 238.1~s from the observation start time.
The intensity in each panel was normalized by the average count rates of the same 
energy band.
The pulse phase has been repeated over two cycles.
A sine curve plus a constant ($ dotted\ curve$) was fitted to the profiles
($plus\ signs$). 
}
\label{pulse}
\end{figure}

To see how the profiles vary with energy, we investigate the energy-resolved profiles
obtained in three different energy bands, 0.2$-$1.2~keV, 1.2$-$3.0~keV, 
and 3.0$-$10~keV.
For this analysis, we focus on the PN dataset as it provides the superior photon statistic in each of the 
considered energy bands.  
Figure~\ref{pulse} shows the profiles (plus signs), where the backgrounds are not subtracted (the 
extracted backgrounds from a photon deficit region in the same CCD chip are very small
and they are at most $\sim 2$\% level compared to their source count rates).
We fit the sinusoidal model to the profiles with allowing all the parameters to be 
free and obtain the following results from the best-fit parameters. Goodness of the
model-fit is given in each panel of the Figure~\ref{pulse}. 
These profiles show no significant phase shift in the chosen energy bands.
The pulse amplitude $2A$ (cf. Equation 1) is found to increase with increasing energy: $0.16\pm0.03$
(0.2 - 1.2 keV), $0.34\pm0.04$ (1.2 - 3.0 keV), and $0.46\pm0.06$ (3.0 - 10 keV).
In the other words, a pulsed (or modulation) fraction (i.e. $A/C$) increases from $8\%\pm2\%$ to 
$23\%\pm3\%$ with increasing energy.
All the quoted errors for the temporal results are at the 90\% confidence level. 

\subsection{Analysis of Optical Monitor (OM) data}

Within 10\arcsec\ from the X-ray position of \AX, we have detected two UV sources in the OM data. 
One UV source (hereafter U1) has found to have the magnitudes of m$_{UVW1}$=15.99$\pm$0.02 and
m$_{UVM2}$=16.49$\pm$0.03 with the corresponding significances of 68.31$\sigma$ and 53.46$\sigma$ 
respectively. On the other hand, the other source (hereafter U2) has the magnitudes of 
m$_{UVW1}$=15.98$\pm$0.03 and m$_{UVM2}$=18.86$\pm$0.26  with the corresponding significances of 
43.81$\sigma$ and 6.70$\sigma$ respectively. 
The corrected count rate of U1 and U2 for both filters are c$_{1}$=3.06$\pm$ 0.06 cts/s (UVW1), 
c$_{1}$=0.82$\pm$0.03 cts/s (UVM2) and c$_{2}$=1.92$\pm$ 0.05 cts/s (UVW1), c$_{2}$=0.06$\pm$0.01 cts/s (UVM2) 
respectively. We subsequently use these count rates to calculate the UV fluxes. 
For the conversion factors in the UVW1 and UVM2 bandpass of 
4.76$\times$10$^{-16}$ erg cm$^{-2}$cts$^{-1}$\AA$^{-1}$ and 
2.20$\times$10$^{-15}$ erg cm$^{-2}$cts$^{-1}$\AA$^{-1}$, we obtain the  
fluxes of (4.36$\pm$0.06)$\times$ 10$^{-12}$ erg cm$^{-2}$ s$^{-1}$ 
and (4.16$\pm$ 0.03)$\times$10$^{-12}$ erg cm$^{-2}$ s$^{-1}$ for U1 in the corresponding filters. 
For U2, the fluxes are found to be (2.66$\pm$0.05)$\times$ 10$^{-12}$ erg cm$^{-2}$ s$^{-1}$ and 
(0.30$\pm$0.01)$\times$ 10$^{-12}$ erg cm$^{-2}$ s$^{-1}$ in UVW1 and UVM2 respectively.

Of these two detected UV sources, U1 is found to have its position coincided with the nominal X-ray position 
of \AX. Furthermore, while there is no optical counterpart can be found for U2, we have identified an optical 
source in $B-$band at the position of U1 (see \S4). The X-ray-to-UV flux ratio is found to have the 
same order-of-magnitude  
as the X-ray-to-optical flux ratio. In view of these properties, we suggest U1 is more likely to be the counterpart 
of \AX.

\section{ DISCUSSION \& SUMMARY }

In this paper, we present a detailed spectro-imaging and temporal analysis of an archival XMM-Newton 
observation of \AX, which has its emission nature not yet been identified unambiguously.

We found that the energy spectrum of \AX\ obtained by the EPIC can be well-described by an absorbed 
multi-temperature plasma model. The inferred column density is 
$n_{H}\sim5\times10^{20}$~cm$^{-2}$ which is far lower than the total Galactic neutral 
hydrogen absorption, $n_{H}\sim10^{22}$~cm$^{-2}$, in the direction towards \AX\ (Kalberla et al. 2005; 
Dickey \& Lockman 1990). This leads us to rule out the extragalactic origin of \AX. 

The unabsorbed X-ray flux inferred from the best-fit model is found to be 
$f_{X}\sim6\times10^{-12}$~ergs~cm$^{-2}$~s$^{-1}$.  
Together with the UV counterpart identified in OM, the inferred X-ray-to-UV flux ratio is $f_{X}/f_{UV}\sim1.5$.
To further constrain the source nature, we also search for any optical identification of \AX\ in the United 
States Naval Observatory (USNO)-B1.0 catalogue (Monet et al. 2003). Within the $20\sigma$ X-ray positional 
uncertainty of \AX\ (see \S3.1), we have identified one optical counterpart with 
$B=17.11$. The position of this source is also found to be consistent with that of U1. 
By using the $n_{H}$ inferred from the X-ray analysis to estimate the foreground extinction, we 
calculate the extinction-corrected optical flux in $B-$band as 
$f_{B}=1.8\times10^{-12}$~ergs~cm$^{-2}$~s$^{-1}$ which implies $f_{X}/f_{B}\sim3.8$.
Such low value of X-ray-to-optical flux ratio can safely 
rule out the possibility as an isolated neutron star which typically has
$f_{X}/f_{B}>10^{3}$ (cf. Haberl 2007).

Through our independent search for the periodic X-ray signal from \AX, we have identified a peak at 
$\sim238.1$~s in the power density spectrum. This has confirmed the periodicity firstly reported by  
Muno et al. (2008). Nevertheless, different from the result reported by Muno et al. (2008), we do not 
find any convincing phase shift in our phase analysis. In any 
case, it is unlikely that this signal with such long period comes from rotation-powered pulsars 
(cf. Manchester et al. 2005).   
Instead, we speculate this periodic signal is possibly from 
a white dwarf. By comparing the uncovered period with the typical values of accreting white dwarf binaries, 
we further suggest that it is likely to be the spin period of the white dwarf and does not 
correspond to an orbital period\footnote{http://asd.gsfc.nasa.gov/Koji.Mukai/iphome/catalog/alpha.html}. 
This interpretation is further supported by the spectral properties. The X-ray spectrum of 
\AX\ can be described by a three-temperature plasma model and clearly shows the presence of iron lines,  
which have often been seen in one type of accreting white dwarf binaries, namely the 
intermediate polars (IPs) (e.g. Patterson 1994; Cropper et al. 2002).
The additional frequencies in the X-ray power spectra which corresponds to orbital or the beat period is absent 
in \AX. In the context of an IP interpretation, 
this suggests that material accretes onto the pole via a disk.

Although the spin period and the spectral properties strongly favor the IP interpretation, 
the observed property of monotonically increasing modulation with increasing energy in 
the light curves requires a further discussion.
In contrast to \AX, many IPs show an opposite trend of decreasing modulation with energy 
(e.g. Norton \& Watson 1989) which is generally explained by the variation of photoelectric absorption 
in the accretion curtain across the observer's line of sight (Rosen, Mason, \& C\'{o}rdova 1988). 
The accretion rate is expected to have the maximum from the sector of the disc that is closest to the 
magnetic pole and fall off with deviation from this sector. This gives rise to a continuous variation of column 
density along the cross-section of the curtain and hence results in the modulation by photoelectric absorption. 
The minimum of the light curve is thus at the phase when the accreting 
pole, where the absorption is the largest, is pointing toward the observer. As the effect of photoelectric absorption 
is more prominent in the soft band than in the hard band, a trend of decreasing pulsed fraction with increasing photon 
energy is not unexpected in this scenario (cf. Figure~9 in Rosen et al. 1988). 

Although the behaviour of decreasing amplitude with increasing energy  
has been observed in many IPs, a number of them do deviate from this general trend. 
For example, a clear increase in amplitude modulation with increasing
energy is also found in V2306 Cygni (WGA J1958.2+3232) 
by the ASCA observation (Norton et al. 2002). 
Norton \& Mukai (2007) found that an IP candidate XY Ari also 
shows an increasing modulation with increasing energy in the XMM-Newton data, whereas 
it shows a decreasing behaviour with increasing 
energy in the RXTE data (see Figure 1 and Figure 5 in Norton \& Mukai 2007). 
Another IP that shows a modulation different from the general trend 
is IGR J00234+6141 in the RXTE observations 
(see Table 2 in Butters et al. 2011). However, the XMM-Newton/EPIC-PN 
observation of this object shows no modulation above 2 keV 
(Anzolin et al. 2009). There is another IP, PQ Gem, which shows an increasing modulation 
in sub-divided low energy bands and a decreasing modulation in high energy bands of 
ASCA and RXTE observations (James et al. 2002). On the other hand, a few other IPs show 
a constant modulation in RXTE observations (e.g. Butters et al. 2007, 2008) and
some show no amplitude modulation in the EPIC-PN energy band (e.g. de Martino et al. 2005). 
Since a diversity of temporal behaviour has been observed, 
energy dependency of amplitude modulation is a weaker criterion to constrain the property of an IP. 

To further probe the temporal behaviour  
of \AX, we have also analysed the archival RXTE observations for this source 
(ObsIDs: 90070-04-01-00, 90070-04-01-01, 90070-04-01-02, 90070-04-02-00, 90070-04-02-01 and 90070-04-02-02). 
However, no pulsed signal from \AX\ was detected from all these datasets. 
On the other hand, we found that the emission line features at 
$\sim6.5$~keV in the RXTE PCA spectrum. Hence, we speculate that the 
non-detection of pulsation in the RXTE data can be ascribed to the lack of imaging capability in resolving the source 
in a crowded region. The upcoming missions with focusing optics for hard X-rays, such as NuSTAR (Harrison et al. 2010), 
will certainly provide the appropriate instruments for further investigation of this interesting IP candidate.

The X-ray spectra of many IPs are described by a complex absorption model as expected in the context of 
accretion curtain (e.g. Ramsay et al. 2008). However, such additional absorption component is not required in modeling 
the spectrum of \AX.  
The lack of complex and high absorption in \AX\ might 
indicate a low inclination angle such that the observed X-rays are not passing through the accretion curtains.
On the other hand, the observed hydrogen column density
(i.e. $n_{\rm H}\sim5\times10^{20}$~cm$^{-2}$) is much lower than typically observed in other IPs
(cf. Ezuka \& Ishida 1999). The absence of high values of absorption can possibly due to the low accretion rate.
Such low absorption have also been seen in several cases, such as HT Cam (de Martino et al. 2005),
EX HYa (Mukai et al. 2003) and V1025 Cen (Hellier et al. 1998) along with a few other IP sources in which
a strong soft component was always prominent (Evan \& Hellier 2007). Nevertheless, such soft component is not observed in
the case of \AX. In the scenario of a low inclination angle, the missing soft component may simply due to the 
geometrical effect (Evan \& Hellier 2007). Alternatively, this component might be much softer than the spectral 
coverage of XMM-Newton EPIC so that it is not revealed in this observation.

The spectral results of \AX\ suggest a relatively low shock temperature of $\sim11$~keV (cf. Table 1). 
Very recently, Yuasa et al. (2010) have investigated the shock temperatures of 
15 IPs and found two sources also with low temperatures, FO Aqr ($\sim$ 14 keV) and 
EX Hya ($\sim$ 12 keV).
In another systematic study of 23 IPs,  
the lowest temperature is found to be $\sim$ 12 keV for the source DO Dra (Brunschweiger et al. 2009). 
Furthermore, we notice that the temperature of a new IP RX~J0704 has exhibited a change from a high value of 
$>44$~keV to $\sim$11~keV in eight months (Anzolin et al. 2008). 
There is an IP, AE Aqr, which shows exceptionally low temperatures plasma of 0.1 - 7 keV
(e.g. Choi et al. 1999; Choi \& Dotani 2006; Mauche 2009). However, for AE Aqr, it is widely believed that
most of the transferred material from the companion does not reach the magnetic poles due to a magnetic
propeller effect (e.g., Wynn et al. 1997; Choi \& Dotani 2006).
Based on aforementioned temperature distribution in IPs, it is clear that the observed temperature 
of \AX\ is not significantly different from a few other IPs and also the observed low plasma temperature 
may not be persistent. A frequent monitoring of this source is encouraged.

In summary, based on our temporal and spectral analysis, we suggest that \AX\ belongs to the 
IP sub-classification of cataclysmic variables. 
We would like to stress that, including XY Ari, there are only 8 IPs/IP candidates which show spin period less than 240~s. 
Owing to this small population, 
the exact selection criteria to classify a source as an IP is not well-defined for this short-period sub-class, 
which so far can only be constrained on the 
basis of the six observational criteria given by Patterson (1994). It is not certain if all 
six of these characteristics should be exhibited by an IP. For a further investigation of this new IP, 
a multi-wavelength observation campaign is required to unveil the physical and geometrical 
configuration of this system. In particular, the determination of the orbital period of this source through 
a dedicated optical observation will definitely play a key role in determining the system parameters such as orbital 
inclination. Furthermore, a dedicated series of X-ray observations is desirable to search for the possible 
eclipses from \AX. Through the timing of the eclipse ingress and egress, one is able to determine the size of 
the X-ray emitting region and hence put an additional observational constraint on the emission
nature of this system.

\section*{Acknowledgments}
The authors would like to thank the anonymous referee for the useful comments.  
CYH is supported by the research fund of Chungnam National University in 2011. 


\clearpage
\begin{landscape}
\begin{table}
\begin{center}
\caption{Best-fit spectral parameters of \AX.}
\end{center}
\resizebox{!}{3.3cm}{
\begin{tabular}{lccccc}
\hline\hline
 & MEKAL & MEKAL+MEKAL & MEKAL+MEKAL & MEKAL+MEKAL+MEKAL & PL \\\hline\hline
$n_{H}$ ($10^{20}$~cm$^{-2}$) & $4.60\pm0.31$ & $4.74^{+0.34}_{-0.35}$ & $5.67^{+0.42}_{-0.40}$ & 
 $5.46^{+0.40}_{-0.38}$ & $12.08^{+0.57}_{-0.56}$\\
$kT_{1}$ (keV) & $6.42\pm0.23$ & $1.00^{+0.04}_{-0.3}$ & $0.81\pm0.03$  & $0.65\pm0.04$ & - \\
$kT_{2}$ (keV) & -  & $8.34^{+0.40}_{-0.42}$ & $7.75^{+0.44}_{-0.65}$  & $1.78^{+0.39}_{-0.15}$ & - \\
$kT_{3}$ (keV) & -  & - & - & $10.84^{+1.40}_{-0.96}$ & - \\
Fe $^{\rm a}$ & - & 1.0 (fixed) & $0.72\pm0.08$ & 1.0 (fixed) & - \\
$\Gamma$  & - & - & - & - & $1.84\pm0.02$ \\
Norm$_{1}$ $^{\rm b}$ & $(2.92\pm0.03)\times10^{-3}$ & $(1.46^{+0.18}_{-0.16})\times10^{-4}$ & 
$(1.55^{+0.22}_{-0.20})\times10^{-4}$ & $(8.56^{+1.23}_{-0.87})\times10^{-5}$ & - \\
Norm$_{2}$ $^{\rm b}$ & - & $(2.70^{+0.03}_{-0.04})\times10^{-3}$ & 
$(2.78\pm0.04)\times10^{-3}$ & $(3.56^{+1.56}_{-0.75})\times10^{-4}$ & - \\
Norm$_{3}$ $^{\rm b}$  & - & - & - & $(2.46^{+0.14}_{-0.07})\times10^{-3}$ & - \\
Norm$_{\rm PL}$ $^{\rm c}$  & - & - & - & - & $(1.07\pm0.02)\times10^{-3}$ \\\hline
$\chi^{2}$ & 1099.53 & 737.50 & 698.87 & 621.20 & 1003.51 \\
D.O.F. & 583 & 581 & 580 & 579 & 583 \\
\hline
\end{tabular}}
\\
\\
$^{\rm a}${\footnotesize The abundance of iron relative to the solar photospheric values.}\\
$^{\rm b}${\footnotesize The normalization of MEKAL model is expressed as 
$(10^{-14}/4\pi D^{2})\int N_{e}N_{H}dV$ where $D$ is the source distance 
in cm and $N_{e}$ and $N_{\rm H}$ are the electron and hydrogen densities in cm$^{-3}$.}\\
$^{\rm c}${\footnotesize The normalization of the power-law model (PL) 
at 1~keV in units of photons~keV$^{-1}$~cm$^{-2}$~s$^{-1}$.} 
\end{table}
\end{landscape}

\bsp

\label{lastpage}
\end{document}